\documentclass[aps,showpacs,preprintnumbers,amsmath, amssymb]{revtex4}

\usepackage{float}
\usepackage{graphics,epsfig}
\usepackage{graphicx}
\usepackage{dcolumn}
\usepackage{bm}

\begin{document}
\baselineskip=0.8 cm
\title{{\bf Influence on the entropic force by the virtual degree of freedom on the holographic screen}}

\author{Qiyuan Pan,$^{1,2}$ Bin Wang$^{1}$}
\affiliation{$^{1}$Department of Physics, Fudan University, 200433
Shanghai, China \\
$^{2}$Institute of Physics and Department of Physics, Hunan Normal
University, 410081 Changsha, China } \vspace*{0.2cm}
\begin{abstract}
\baselineskip=0.6 cm
\begin{center}
{\bf Abstract}
\end{center}

We generalize the study of entropic force to a general static
spherical spacetime and examine the acceleration, temperature,
equation of gravity and the energy associated with the holographic
screen in this general background. We show that the virtual degree
of freedom on the holographic screen plays a crucial role in
interpreting field equations of gravity based on thermodynamical
perspective.

\end{abstract}

\maketitle
\newpage

\section{Introduction}
The discovery of the thermodynamical properties of black holes
inspired deep thinkings on the profound relation between gravity and
thermodynamics. A pioneer work on this respect was done by Jacobson
who disclosed that the gravitational Einstein equation can be
derived from the relation between the horizon area and entropy,
together with the Clausius relation $\delta Q=TdS$ \cite{Jacobson}.
This derivation suggests that the Einstein equation for the
spacetime metric has a predisposition to thermodynamic behavior.
Jacobson's investigation has been extended to the gravitational
theory beyond Einstein gravity, including $f(R)$ gravity \cite{2},
the Gauss-Bonnet gravity, the scalar tensor gravity and more general
Lovelock gravity \cite{3,4}. The study has also been generalized to
the cosmological context \cite{5}, including the general braneworld
models \cite{6}. For a review, see \cite{10}.

Recently a constructive new idea on the relation between gravity and
thermodynamics was proposed by Verlinde \cite{Verlinde}. He argued
that gravitational interaction is not fundamental and gravity is an
emergent entropic force originated from the change of information
associated with the positions of bodies of matter. Motivated by
Bekenstein's entropy bound, Verlinde postulated that when a test
particle with mass $m$ approaches a holographic screen from a
distance $\Delta x$, the change of entropy on the holographic screen
is
\begin{eqnarray}\label{change-entropy}
\Delta S=2\pi k_{B}\frac{mc}{\hbar}\Delta x.
\end{eqnarray}
The entropic force can arise in the direction of increasing entropy
and is proportional to the temperature, $F=T\Delta S/\Delta x$.
Adopting the equipartition law of energy
\begin{eqnarray}
E=Mc^{2}=\frac{1}{2}Nk_{B}T,
\end{eqnarray}
where $M$ represents the mass enclosed by the holographic screen and
$N$ is the degree of freedom associated with the screen, one can
determine the expression for the temperature $T$. Assuming the total
number of bits $N$ on the holographic screen being proportional to
the area of the screen $A$, $N=\frac{Ac^{3}}{G\hbar}$, Verlinde got
the Newton's law of gravity
\begin{eqnarray}\label{Newton-gravitation}
F=G\frac{Mm}{r^{2}},
\end{eqnarray}
where he inserted $A=4\pi r^2$. Similar observation of the Newton's law of gravity was also obtained by combining the thermodynamical relation
$S=E/2T$ with the equipartition law of energy for the horizon degree of freedom \cite{Padmanabhan-2009}. The derivation has been further extended
to the relativistic situation and the Einstein equation describing the law of gravity in the relativistic case was obtained \cite{Verlinde}. The
remarkable idea of the entropic force has been heated discussed recently, see for example \cite{Shu-Gong}-\cite{Liu}.

The assumption that the total number of bits in proportional to the
area $A$ of the holographic screen is crucial in Verlinde's
derivation. This assumption looks reasonable by taking account of
the holographic principle originated from the `It from bit' picture
of Wheeler \cite{Wheeler}. However it was argued in
\cite{Kaul-Majumdar,Majumdar} that a variant of this picture that
takes better account of the symmetries of general relativity is
shown to yield corrections to the counting of the degree of freedom
that are logarithmic in the area, with a finite, fixed coefficient.
If we insert this virtual degree of freedom in Verlinde's
derivation, it would be interesting to ask what kind of influence
the virtual degree of freedom will bring to the field equations of
gravity. In this work we will try to examine this influence.

We will start our discussion with a general class of spherically symmetric and static spacetime and apply Verlinde's approach to investigate the
acceleration, temperature and energy on holographic screens and test the entropic force in this general background. In the general spherically
symmetric case with a horizon, we will show that the results are consistent with that of Verlinde. This serves the main purpose in this work.
Further, we will examine the influence on the Newton gravity and Einstein equation caused by the virtual degree of freedom on the holographic
screen.

\section{Emergent gravity in general static spherically symmetric spacetime}

We consider a general class of spherically symmetric and static four-dimensional spacetime
\begin{eqnarray}\label{metric}
ds^2=-f(r)dt^{2}+\frac{1}{h(r)}dr^{2}+R^{2}(r)(d\theta^{2}+\sin^{2}\theta
d\varphi^{2}),
\end{eqnarray}
where we relax the condition $f(r)=h(r)$, which is accidentally
verified in four dimensions but, in fact, there is no reason for it
to continue to be valid in the general scenario. In this spirit,
black hole and wormhole solutions as well as star solutions on the
brane have been obtained in the last years \cite{aa}. For this
general class of spherically symmetric line element $f(r)$ and
$h(r)$ vanish at the black hole event horizon $r=r_+$, that is
$f(r)=f'(r_{+})(r-r_{+})$ and $h(r)=h'(r_{+})(r-r_{+})$ as
$r\rightarrow r_{+}$, but not necessarily at the same rate. To
ensure the metric to be asymptotically flat, we require
$f(r)|_{r\rightarrow\infty}=h(r)|_{r\rightarrow\infty}=1$. We have
adopted units $G=c=\hbar=\kappa_{B}=1$. The surface gravity of the
event horizon can be determined by
\begin{equation}\label{surface gravity}
\kappa=\frac{1}{2}\sqrt{f'(r_{+})h'(r_{+})},
\end{equation}
where the prime denotes the derivative with respect to $r$.

For the general static spherically symmetric spacetime, we can
express the timelike Killing vector as \cite{Verlinde,Liu,Konoplya}
\begin{equation}\label{Killing}
\xi_{\mu}=(-f(r), 0, 0, 0).
\end{equation}
The generalization of the Newton's potential is
\begin{equation}\label{potential}
\phi=\frac{1}{2}\ln (-g^{\mu\nu}\xi_{\mu}\xi_{\nu})
=\frac{1}{2}\ln[f(r)],
\end{equation}
which can be reduced to the ordinary Newtonian potential. This can
easily be seen if we look at the simplest case, the Schwarzschild
black hole with $f(r)=h(r)=1-2M/r$ , we can obtain $\phi=-M/r$  in
the large $r$  limit, which is just the ordinary Newtonian
potential.

The potential can be used to define the foliation of the space and
the holographic screen is at the surface of constant redshift. The
acceleration perpendicular to the screen can be expressed as
\begin{equation}\label{acceleration}
a^{\mu}=-\nabla^{\mu}\phi=\left(0,-\frac{h(r)f'(r)}{2f(r)},0,0\right).
\end{equation}

The local temperature on the screen can now be defined by
\cite{Verlinde}
\begin{equation}\label{temperature}
T=\frac{1}{2 \pi}e^{\phi}\sqrt{\nabla^{\mu}\phi\nabla_{\mu}\phi} =
\frac{1}{4\pi}\sqrt{\frac{h(r)}{f(r)}}f'(r),
\end{equation}
where a redshift factor $e^{\phi}$ was inserted because the
temperature $T$ is measured with respect to the reference point at
infinity. This temperature can be considered as Unruh temperature
which is closely related to the local acceleration perpendicular to
the screen. As Unruh showed, an observer in an accelerated frame can
experience this temperature. Furthermore in the spirit of
\cite{Verlinde} it was argued that this temperature is actually
required to cause an acceleration equal to $a$.

At the black hole event horizon $r=r_{+}$, we have
\begin{equation}\label{temperature horizon}
T|_{r_{+}}=\frac{1}{4\pi}\sqrt{f'(r_{+})h'(r_{+})}=\frac{\kappa}{2\pi},
\end{equation}
which is just the Hawking temperature $T_{H}$ relating to the
surface gravity $\kappa$. When $r\gg r_{+}$, the metric coefficients
can be expanded in power series of $1/r$ in the form
\begin{equation}\label{expansionF}
f(r)=1-\frac{2M}{r}+\frac{Q^{2}}{r^{2}}+O\left(\frac{1}{r^{3}}\right),
\end{equation}
\begin{equation}\label{expansionH}
h(r)=1-\frac{2M'}{r}+\frac{Q'^{2}}{r^{2}}+O\left(\frac{1}{r^{3}}\right),
\end{equation}
where $M, M', Q, Q'$ are expansion parameters. When it is far away
from the black hole horizon, we have
\begin{equation}
T|_{\infty}=0.
\end{equation}
The local temperature on the holographic screen is measured with
respect to the reference point at infinity. When the holographic
screen coincides with the black hole horizon, the temperature on the
screen is just the black hole Hawking temperature. When the screen
moves away from the horizon, the temperature on the screen  changes
until it vanishes at spatial infinity.

The force on a particle located very close to the holographic screen
can be calculated as
\begin{equation}\label{entropic force}
F^{\mu}=T\nabla^{\mu}S=-m
e^{\phi}\nabla^{\mu}\phi=\left(0,-\frac{mh(r)f'(r)}{2\sqrt{f(r)}},0,0\right),
\end{equation}
where $m$ is the mass of the test particle. Obviously, with the help
of Eq. (\ref{acceleration}), $F^{\mu}=m e^{\phi}a^{\mu}$ is the
second law of Newton. This is the gravitational force required to
keep a particle at fixed position near the screen as measured from a
reference point at infinity \cite{Verlinde}. The exponent $e^{\phi}$
is the redshift factor, which was inserted in $T$ in Eq.
(\ref{temperature}), because the temperature $T$ is a local quantity
which is measured with respect to the reference point at infinity.
The magnitude of this force is
\begin{equation}
F=\sqrt{F^{\mu}F_{\mu}}=\frac{m}{2}\sqrt{\frac{h(r)}{f(r)}}f'(r).
\end{equation}
At the black hole event horizon, it is interesting to note that
\begin{equation}
F|_{r_{+}}=\frac{1}{2}m\sqrt{f'(r_{+})h'(r_{+})}=m\kappa.
\end{equation}
When $r\gg r_{+}$, using the expansions of the metric coefficients,
we arrive at the Newton's law of gravity
\begin{equation}
F=\frac{M_{0}m}{r^{2}},
\end{equation}
with the reduced mass
\begin{equation}\label{1}
M_{0}=M\left[1+(M-M')\frac{1}{r}\right]-\frac{Q^{2}}{r}.
\end{equation}
Since we have neglected terms of order $O(1/r^{2})$ and higher, $Q'$
does not appear here. The reduced mass $M_{0}$ enclosed by the
holographic screen in the general static spherical black hole
spacetime returns to the black hole mass $M_0=M$ and
$M_{0}=M-Q^{2}/r$ for Schwarzschild black hole and the
Reissner-Nordstr\"{o}m black hole respectively as observed in
\cite{Liu}.

Now we apply the equipartition relation to calculate the energy on
the holographic screen. Adopting the assumption $N=A$, we have
\begin{equation}\label{energy}
E=\frac{1}{2}\int_{\mathbb{S}}TdN=\frac{1}{4\pi}\int_{\mathbb{S}}e^{\phi}\nabla\phi
dA=\frac{r^{2}}{2}\sqrt{\frac{h(r)}{f(r)}}f'(r).
\end{equation}
When the screen coincides with the black hole horizon, we have
\begin{equation}\label{energy horizon}
E|_{r_{+}}=\frac{1}{2}r_{+}^{2}\sqrt{f'(r_{+})h'(r_{+})}=r_{+}^{2}\kappa,
\end{equation}
which depends on the surface gravity $\kappa$. Considering at the
horizon $S=A/4=\pi r_+^2, T|_{r_+}=\kappa/2\pi$, we have $E=2TS$,
which was first obtained in \cite{Padmanabhan-2009} in
reinterpreting the law of equipartition from the thermodynamic
description at the horizon. When $r\gg r_{+}$, neglecting terms of
order $O(1/r^{2})$ and the higher, we get
\begin{equation}
E=M\left[1+(M-M')\frac{1}{r}\right]-\frac{Q^{2}}{r},
\end{equation}
which is just the reduced mass $M_{0}$ related to the black hole in
Eq. (\ref{1}). If we employ the idea in the holographic principle
which asserts that the maximum possible number of degrees of freedom
is given by a quarter of the area of the screen $S=A/4=\pi r^{2}$,
we again can have $E=2TS$ from Eqs. (\ref{temperature}) and
(\ref{energy}). This shows that the finding at the horizon in
\cite{Padmanabhan-2009} keeps on all holographic screens at constant
redshifts. On any screen, the law of equipartition is equivalent to
the thermodynamic description.

\section{Virtual degree of freedom influence on the emergent gravity}

In the above section, we have discussed the entropic force for the
general static spherical spacetime. In getting the entropic force we
have adopted the assumption that the number of bits $N$ on the
holographic screen is proportional to the area of the screen $A$.
This assumption can be understood from Wheeler's `It from Bit'
picture \cite{Wheeler}. Consider that there are $p$ numbers of two
dimensional finite `floating lattice' with size of a Planck area
$l_p^2$ covering the holographic screen. Macroscopically, the
classical area of the screen $A\gg l_p^2 $. Assume that binary
variables (bits) are distributed randomly on the lattice and
typically the variables could be with elementary spin $1/2$ of
$SU(2)$ group \cite{Kaul-Majumdar,Majumdar}. The Hilbert space of
quantum states defined by these spin $1/2$ variables has a
dimensionality ${\cal N}(p)= 2^{p}$ which leads to the number of
degree of freedom characterizing the holographic screen $N=p\ln 2$.
When $p\gg 1$, the lattices can be taken to approximate the
macroscopic screen, $A/l_p^2=\xi p$. For a choice $\xi=\ln 2$, one
has $N=A/G$, where we still keeps $c=\hbar=\kappa_{B}=1$.

The generality of the above scenario in counting the degree of
freedom of the holographic screen makes it appealing also to the
quantum consideration. However in quantum aspect one has to consider
the symmetry, which is a crucial aspect of any quantum approach.
Considering the elementary variables are spin $1/2$ variables under
spatial rotations, the symmetry criterion on the physical Hilbert
space requires that the Hilbert space consists of states which are
compositions of elementary $SU(2)$ doublet states with vanishing
total spin. This is the natural choice of the symmetry since in the
`It from Bit' picture, the basic variables are spin 1/2 variables.
This symmetry was also shown arise naturally in the non-perturbative
quantum general relativity approach known as quantum geometry
\cite{Kaul-Majumdar,Majumdar}. In the large $p$ limit, it was shown
that the dimensionality of physical Hilbert space
\cite{Kaul-Majumdar,Majumdar}
\begin{eqnarray}
dim{\cal H_{S}}\equiv{\cal N}(p)\approx\frac{2^{p}}{p^{3/2}} ,
\end{eqnarray}
and the number of bits on the holographic screen becomes
\begin{eqnarray}\label{corrected bits}
N=\frac{A}{G}\left[1-
\frac{3}{2}\left(\ln\frac{A}{4G}\right)/\left(\frac{A}{4G}\right)\right],
\end{eqnarray}
where the overcounting on the degree of freedom has been taken out
by considering the symmetry in the quantum approach.

It would be interesting to examine how the virtual degree of freedom
on the holographic screen when the quantum aspect is taken into
account will influence the equation of gravity. Employing the
equipartition law of energy, we can obtain the entropic force due to
the change of the virtual information on the screen
\begin{eqnarray}\label{corrected entropic force}
F=T\frac{\Delta S}{\Delta
x}=\frac{GMm}{r^{2}}\frac{1}{1-\frac{3}{2}\left(\ln\frac{A}{4G}\right)/\left(\frac{A}{4G}\right)}.
\end{eqnarray}
Comparing with Eq. (\ref{Newton-gravitation}),
we clearly see that the virtual degree of freedom  brings the
quantum correction to the entropic force. Macroscopically, when we
consider $A/G\gg1$ and
$\left(\ln\frac{A}{4G}\right)/\left(\frac{A}{4G}\right)\ll1$, this
quantum correction will be neglected.

In reference \cite{Modesto}, the authors assumed that on the surface
the information scales proportional to the area of the surface
$N=A/G$, and they got the modified force
$F=\frac{GMm}{r^2}[1+4G\frac{\partial S}{\partial A}]_{A=4\pi r^2}$.
Macroscopically, since $\left(\frac{\partial S}{\partial
A}\right)_{A=4\pi r^2}\ll 1$, this quantum correction can be
neglected. In our study we considered the virtual degree of freedom
on the holographic screen (23), which leads to our (24).  Although
macroscopically in Eq. (24) the quantum correction can be neglected
as well, microscopically our result is different from that in
\cite{Modesto}, since we considered the virtual information on the
screen.

We can further consider the influence on the Einstein equation
caused by the virtual degree of freedom on the holographic screen.
From Eq. (\ref{corrected bits}), it is easy to arrive at
\begin{eqnarray}\label{corrected dN}
dN=\left(\frac{1}{G}-\frac{6}{A}\right)dA.
\end{eqnarray}
Expressing the energy in terms of the total enclosed mass $M$ and
employing the law of equipartition, we have
\begin{eqnarray}\label{Corrected-MTdN}
M=\frac{1}{2}\int_{\mathbb{S}}TdN=\frac{1}{4\pi G}\int_{\mathbb{S}}
\left(1-\frac{6G}{A}\right)e^{\phi}\nabla\phi dA.
\end{eqnarray}
Following the same logic in \cite{Verlinde}, we can get the integral
relation
\begin{eqnarray}\label{corrected Ein-Eq}
2\int_{\Sigma}
(T_{\mu\nu}-\frac{1}{2}Tg_{\mu\nu})N^{\mu}\xi^{\nu}dV=\frac{1}{4\pi
G}\int_{\Sigma}R_{\mu\nu}N^{\mu}\xi^{\nu}dV-\frac{3}{2\pi}\int_{\mathbb{S}}
\frac{e^{\phi}\nabla\phi}{A} dA.
\end{eqnarray}
The second term on the right-hand-side is an additional term
compared with Eq. (5.37) derived in \cite{Verlinde}. This term  is
caused by the quantum correction to the virtual degree of freedom on
the holographic screen which  brings a surface correction to the
Einstein equation. We can use the simplest Schwarzschild spacetime as an example to further see the role played by this term. For the Schwarzschild spacetime
\begin{eqnarray}
e^{\phi}\nabla\phi=\frac{1}{2}\sqrt{\frac{h(r)}{f(r)}}f'(r)=\frac{MG}{r^{2}},
\end{eqnarray}
we can work out the integration
\begin{eqnarray}
\int_{\mathbb{S}} \frac{e^{\phi}\nabla\phi}{A} dA=4\pi
\int_{\mathbb{S}}\frac{GM}{4\pi r^{2}}\frac{dA}{A}=-\frac{4\pi
MG}{A}.
\end{eqnarray}
Now, the integral relation (27) becomes
\begin{eqnarray}\label{integral relation}
2\int_{\Sigma}
(T_{\mu\nu}-\frac{1}{2}Tg_{\mu\nu})N^{\mu}\xi^{\nu}dV=\frac{1}{4\pi
G}\int_{\Sigma}R_{\mu\nu}N^{\mu}\xi^{\nu}dV+2\int_{\Sigma}
\epsilon(T_{\mu\nu}-\frac{1}{2}Tg_{\mu\nu})N^{\mu}\xi^{\nu}dV,
\end{eqnarray}
where we have used $M=\int_{\Sigma}
(T_{\mu\nu}-\frac{1}{2}Tg_{\mu\nu})N^{\mu}\xi^{\nu}dV$ and
$\epsilon=6G/A\ll1$ in our discussion. Note that Eq. (\ref{integral
relation}) is always true for arbitrary volume element, so we 
write in the form the correction to the Einstein equation 
\begin{eqnarray}\label{corrected Ein}
R_{\mu\nu}=8\pi G(T_{\mu\nu}-\frac{1}{2}Tg_{\mu\nu})+J_{\mu\nu},
\end{eqnarray}
where $J_{\mu\nu}=8\pi G\epsilon(\frac{1}{2}Tg_{\mu\nu}-T_{\mu\nu})$
is the very small correction which is from the surface term
$\frac{3}{2\pi}\int_{\mathbb{S}} \frac{e^{\phi}\nabla\phi}{A} dA$.

Different from the usual field equations, the term $J_{\mu\nu}$ is a
nonlocal effect \cite{Easson-Smoot,Yi-Fu}, which is determined by
the holographic description of boundary physics in the frame of
holography. The additional surface term brings the similarity to the
surface term in the Einstein equation discussed in
\cite{Easson-Smoot,Yi-Fu}, where it was argued arising from the
surface term in the action. It was claimed that the correction to
the Einstein equation arising from the surface term in the action
can be used to explain the universe acceleration
\cite{Easson-Smoot,Yi-Fu}. In our case, Eq. (\ref{corrected Ein-Eq})
was derived in a general static background with a time like Killing
vector. It would be interesting to examine the influence due to the
virtual degree of freedom on the holographic screen on the
gravitation equation in the dynamical spacetimes and see whether the
virtual information on the screen can result in the acceleration of
the universe.

\section{Conclusions}

In this work we have generalized Verlinde's approach on the entropic
force to a general static spherical spacetime. We have studied the
acceleration, temperature, gravitational equation and the energy
associated with the holographic screen in the general background.
Adopting the assumption that the degree of freedom on the
holographic screen $N=A$, we have got the general reduced mass
enclosed by the holographic screen. For Schwarzschild and
Reissner-Nordstr\"{o}m black holes, our general reduced mass returns
to that discussed in \cite{Liu}. We observed that the relation
$S=E/2T$ which can be reinterpreted as the law of equipartition not
only holds on the black hole horizon as argued in
\cite{Padmanabhan-2009}, but also on all holographic screens.

In \cite{Verlinde}, the degree of freedom on the holographic screen
was assumed in proportional to the area of the screen. This
assumption is crucial in deriving the entropic force. Starting from
the Wheeler's `It from Bit' picture, we have considered the quantum
effect in counting the degree of freedom on the screen. We have
investigated the influence of the virtual degree of freedom on the
equations of gravity. The Newton's law and the Einstein equation are
both modified due to the quantum effect in the virtual degree of
freedom. This shows that the virtual degree of freedom on the
holographic screen plays a crucial role in interpreting field
equations of gravity based on thermodynamical perspective. It would
be interesting to generalize our study to the dynamical spacetime
and examine the effect of the virtual degree of freedom on the
holographic screen on cosmic evolution.

\begin{acknowledgments}

This work was partially supported by the National Natural Science
Foundation of China. Qiyuan Pan was also supported by the China
Postdoctoral Science Foundation.

\end{acknowledgments}

\end{document}